\documentclass[journal]{IEEEtran}

\usepackage{url}
\usepackage{graphicx}
\usepackage{amsmath}
\usepackage{amsfonts}
\usepackage{amssymb, color}
\usepackage[noadjust]{cite}

\newtheorem{theorem}{Theorem}

\begin{document}
\vspace{-5mm}

\title{Optimal Power and Time Allocation for WPCNs with Piece-wise Linear EH Model}
\vspace{-5mm}

\author{Slavche Pejoski, Zoran Hadzi-Velkov, and Robert Schober 
\vspace{-9mm}

\thanks{S. Pejoski and Z. Hadzi-Velkov are with the Faculty of Electrical Engineering and Information Technologies, Ss. Cyril and Methodius University, 1000 Skopje, Macedonia (email: \{slavchep, zoranhv\}@feit.ukim.edu.mk).

R. Schober is with the Institute for Digital Communication, Friedrich-Alexander-University Erlangen-N\"urnberg, Erlangen D-91058, Germany (email: robert.schober@fau.de) 
}
}

\markboth{}{Shell \MakeLowercase{\textit{et al.}}: Bare Demo of IEEEtran.cls for Journals} \maketitle

\vspace{-3mm}
\begin{abstract}
We propose a novel transmission protocol for harvest-then-transmit wireless powered communication networks, which takes into account the non-linearity of the energy harvesting (EH) process at the EH users  and maximizes the sum rate in the uplink.
We assume a piece-wise linear energy harvesting model and provide  expressions for the optimal transmit power of the base station (BS), the duration of the  EH phase, and the duration of the uplink information transmission phases of the users. The obtained solution provides insight regarding the significance of the non-linear EH model on the optimal resource allocation. Simulations unveil the growing impact of the saturation effect, which occurs for high received radio frequency powers, as the average and the maximum instantaneous transmit powers of the BS increase.
\end{abstract}


\vspace{-5mm}

\section{Introduction}
\vspace{-2mm}
Wireless powered communication networks (WPCNs) are a new type of wireless network that combines  
wireless information transmission and far-field radio frequency (RF) charging of the users' batteries. The charging is carried out through RF energy transmission from the base station (BS) to the energy harvesting (EH) users (EHUs). RF-EH
necessitates new resource allocation schemes for wireless networks \cite{lit2,lit21}. One of the key aspects for
proper resource allocation
is the accurate modeling of the EH process at the EHUs.
The WPCN resource allocation schemes developed in the literature so far mostly rely on a linear EH model, which assumes that the power harvested in an EHU's battery ($P_h$) is linearly dependent on the  RF power incident at the EHU's antenna ($P_{in}$), i.e., $P_h = \eta P_{in}$, where $\eta$ is the energy  conversion efficiency and is independent of $P_{in}$. However, the EH characteristic ($P_{in}$ vs. $P_h$) of practical RF-EH circuits is non-linear \cite{le}, \cite{stoo}.
As $P_{in}$ increases, the non-linear EH circuit eventually causes $P_h$ to become saturated. The recent work in \cite{elena} studies resource allocation for a multi-antenna simultaneous wireless information and power transfer (SWIPT) system for a practical non-linear EH model, where the EH characteristic is approximated via curve fitting. A comparison of several nonlinear EH models can be found in \cite{Kang}. The design and resource allocation of WPCNs are affected by the non-linearity of practical EH circuits, and algorithms designed based on the linear EH model can lead to a  significant performance loss due to model mismatch \cite{elena, Kang, elena2, Xiong}. References \cite{elena} and \cite{elena2}
 show that the saturation of practical EH circuits is responsible for this performance loss. Additionally,  incorporating the saturation effect into the EH model appears  particularly important for maximizing the sum rate in the uplink of WPCNs. In particular, \cite[Theorem 1]{lit21}, which assumes the linear EH model, suggests that the BS should transmit with the maximum possible power when the channel conditions are favorable. As a result, in practice, many of the EHUs may be in saturation due to their non-linear EH circuit. Although the EH model in \cite{elena} shows a very good match with measurement data, its application for EH resource allocation design leads to complicated optimization problems, solvable only via iterative numerical algorithms. These algorithms do not reveal the influence of the non-linearity on optimal resource allocation. To address this issue, in this letter, we adopt a simple piece-wise linear approximation of the non-linear EH characteristic and analyze its influence on the optimal power and time allocation of harvest-then-transmit WPCNs. The motivation for using a piece-wise linear EH model is two-fold: first, it is analytically tractable, and second, it captures the saturation behavior of practical EH circuits. The piece-wise linear  EH model has been already employed in the context of outage performance analysis of relay systems \cite{Dong}, \cite{Zhanq} and secure communication \cite{Shafie}, but not in the context of resource allocation.

The novel contributions of this letter are analytical expressions for the optimal BS power allocation and the optimal time sharing parameters that maximize the average uplink network sum rate  for both constrained and unconstrained instantaneous BS transmit powers under the piece-wise linear EH model.

\vspace{-4mm}
\section{System model and Sum Rate Maximization}
\vspace{-2mm}
We consider a WPCN, which consist of a single BS and $K$ EHUs, employing Time Division Multiple Access (TDMA). All nodes are equipped with a single antenna, and operate in the half-duplex mode. The transmission time is divided into $M$ epochs of equal duration $T$, each containing a single TDMA frame. Each TDMA frame is divided into an EH phase, and $K$ information transfer (IT) phases. 
In epoch $i$, the duration of the EH phase is denoted by $\tau_0(i) T$, whereas the duration of the successive IT phases of all EHUs are denoted by $\tau_1(i)T$,$\ldots$, $\tau_k(i)T$,$\dots$, $\tau_K(i)T$, respectively, where $\tau_k(i)$ are time-sharing parameters  which are dimensionless quantities satisfying $0\leq\tau_k(i)\leq 1$ and $\tau_0(i)+\sum_{k=1}^K\tau_k(i)=1$.
We assume that all wireless links exhibit frequency non-selective block fading, i.e., the channel coefficients are constant in each slot but change from one slot to the next, where each slot corresponds to one epoch.
In epoch $i$, the fading power gain of the $BS-EHU_k$ channel is $x'_k(i)$ and we normalize its value by the power of the additive white Gaussian noise (AWGN) at the receiver, $N_0$, to obtain $x_k(i)=x'_k(i)/N_0$ with average value $\Omega_k=\mathbb{E}[x'_k(i)]/N_0$, where $\mathbb{E}[\cdot]$ denotes expectation. For convenience, the downlink and uplink channels are assumed to be reciprocal and the BS is assumed to have full channel state information in each epoch. The instantaneous transmit power of the BS is denoted by $p_0(i)$ and satisfies the average power constraint $P_{avg}$, i.e., $\mathbb{E}[p_0(i)\tau_0(i)]=P_{avg}$. 

\vspace{-5mm}
\subsection{EH model}
\vspace{-2mm}
The EHUs are equipped with rechargeable batteries and employ a harvest-then-transmit mechanism, i.e., they charge their batteries in the EH phase and then transmit information in the corresponding IT phases during which they spend all of their harvested energy.

The EH process can be modelled by a piece-wise linear EH 
characteristic, which implies that the RF-DC (direct current) conversion curve consists of a linear part corresponding to the linear regime of operation, and a constant (saturation) part corresponding to saturated regime of operation. Therefore, the harvested power of EHU $k$ is modeled as:
\vspace{-2mm}
\begin{equation}
P_{h,k}=\begin{cases}
\eta_kP_{in}, &\eta_kP_{in}\leq P_{Hk}\\
P_{Hk}, &\text{otherwise}
\end{cases}
\vspace{-2mm}
\end{equation}
where $P_{Hk}$ is the peak amount of power that can be harvested by EHU $k$ and $\eta_k$ denotes the EH efficiency ($0<\eta_k<1$).
Thus, the harvested energy of EHU $k$ in epoch $i$ is given by
\vspace{-2mm}
\begin{equation} \label{energy_model1}
E_{k}(i) = T\cdot\min\{N_0\eta_kx_k(i)p_0(i)\tau_0(i), P_{Hk}\tau_0(i)\}.
\vspace{-2mm}
\end{equation}
Therefore, its transmit power during the IT phase is 
$P_{k}(i) = E_k(i)/(\tau_k(i)T)$, and its achievable rate is $r_k(i) = \tau_k(i)\log(1 + P_k(i)x_k(i))$. The average achievable rate of EHU $k$ 
is $\overline R_k = \lim_{M\rightarrow\infty}\frac{1}{M}\sum_{i=1}^M r_k(i)$.

\vspace{-4mm}
\subsection{Unconstrained instantaneous BS transmit power}
\vspace{-2mm}
\label{unconstrained}
Here, we aim at maximizing the sum rate in the uplink subject to an average power constraint at the BS:
\begin{eqnarray} \label{OP1}
\notag
\text{{\bf Q$_0$}}\ \hspace{-1mm}:\hspace{-1mm}\ \underset{\tau_{k}(i), \tau_{0}(i), p_0(i)} {\text{Maximize}}\hspace{-3mm} & &\hspace{-3mm}\frac{1}{M}\sum_{i=1}^M\sum_{k=1}^K\tau_{k}(i)\log\left(1+\frac{{x}_{k}(i)}{\tau_k(i)}\right.\\
\hspace{-3mm}& &\hspace{-23mm}\times\left.\min\{N_0\eta_k x_k(i)p_0(i)\tau_0(i), P_{Hk}\tau_0(i)\}\vphantom{\frac{1}{2}}\right)
\end{eqnarray}
\vspace{-3mm}
\hspace{10mm}\text{s.t.}
\vspace{-5mm}
\begin{eqnarray}
\begin{array}{ll}
&\text{C1}: \frac{1}{M}\sum_{i=1}^Mp_0(i)\tau_0(i) \leq P_{avg} \\ \notag
&\text{C2}: \tau_0(i)+\sum_{k=1}^K \tau_k(i) = 1, \forall i \\ \notag
&\text{C3}: 0 \leq p_0(i),\ \forall i. \\ \notag
\end{array}
\end{eqnarray}
\vspace{-7mm}

The optimal solution of problem {\bf Q$_0$} is given in the following theorem\footnote{We note that if the non-linear EH model [5, Eq. (4), (5)] is applied instead of the piece-wise linear model in (1), the resulting sum-rate maximization problem is non-convex, and does not permit an efficient solution. Applying an exhaustive search is computationally infeasible, since we assume $M\rightarrow\infty$.}.
\begin{theorem}\label{teorem1}
In each epoch $i$, we relabel the EHUs such that:
\begin{equation}
\label{assu}
\frac{P_{H1}}{N_0\eta_1x_1(i)}<\frac{P_{H2}}{N_0\eta_2x_2(i)}<...<\frac{P_{HK}}{N_0\eta_Kx_K(i)}.
\end{equation}
Such relabeling does not affect the objective function which only depends on the sum of the rates of all EHUs.

In each epoch $i$, the optimal power allocation at the BS, $p_0(i)$, if greater than 0, inverts the channel of one of the EHUs, the index of which we denote by $s^*$, such that the harvested power of this EHU is exactly on the boundary between the linear and the saturated part of the EH characteristic curve, i.e., the harvested power is $P_{h,s^*} = \eta_{s^*}p_0(i)N_0x_{s^*}(i) = P_{Hs^*}$. 
Index $s^*$ is determined as follows:
\begin{equation}
\label{s_value}
s^*=\left\{1\leq s\leq K| \frac{A_{s-1}(i)}{\lambda}>z_{s-1}\ \text{and}\ \frac{A_s(i)}{\lambda}<z_s\right\}
\end{equation}
where
\vspace{-3mm}
\begin{equation}
\label{eq_A}
A_s(i)=\hspace{-3mm}\sum_{m=s+1}^K\hspace{-3mm}N_0\eta_mx_m^2(i),
\end{equation}
\vspace{-2mm}
\begin{equation}
\label{eq_B}
 B_s(i)=\sum_{n=1}^sP_{Hn}x_n(i),
\end{equation}
\vspace{-2mm}
\begin{equation}
z_s=(B_s(i)-1)/W\left((B_s(i)-1)/e\right),
\end{equation}
$W(\cdot)$ is the Lambert-W function, and $\lambda$ is chosen such that equality holds in C1. Thus, the optimal power allocation at the BS is:
\vspace{-2mm}
\begin{equation}
\label{pow_BS}
p_0(i) = \begin{cases}
P_{Hs^*}/(N_0\eta_{s^*}x_{s^*}(i)), & A_0(i)>\lambda\\
0, & \text{otherwise}
\end{cases}
\vspace{-2mm}
\end{equation}
Therefore, the harvested power of the EHUs with indices $n\in S_2^*=\{1,2,\ldots,s^*-1\}$ are in the saturation regime, i.e., the harvested powers are $P_{Hn}$, respectively, whereas the harvested powers of the EHUs with indices $m\in S_1^*=\{s^*+1,s^*+2,\ldots,K\}$ are in the linear regime, i.e., the harvested powers are $P_{h,m}=\eta_{m}p_0(i)N_0x_{m}(i)$, respectively.

The optimal duration of the IT phase of the EHUs in the linear regime is given by
\vspace{-2mm}
\begin{equation}
\label{taum_EHS}
\tau_m(i) = p_0(i)N_0\eta_mx_m^2(i)\tau_0(i)/(C_{s^*}(i)-1),
\vspace{-2mm}
\end{equation}
and the optimal duration of the IT phase of the EHUs in the saturated regime is given by
\vspace{-2mm}
\begin{equation}
\label{taun_EHS}
\tau_n(i) = P_{Hn}x_n(i)\tau_0(i)/(C_{s^*}(i)-1),
\vspace{-2mm}
\end{equation}
where $\tau_0(i)$ is the optimal duration of the EH phase, given by
\vspace{-2mm}
\begin{equation}
\label{tau_BS}
\tau_0(i) = \left(1+\frac{p_0(i)A_{s^*}(i)+B_{s^*}(i)}{C_{s^*}(i)-1}\right)^{-1}.
\vspace{-2mm}
\end{equation}
Constant $C_{s^*}(i)$ is obtained as the solution to the following transcedent equation:
\vspace{-2mm}
\begin{eqnarray}
\notag
\label{eq_C}
\log(C_{s^*}(i))\hspace{-3mm}&-&\hspace{-3mm}\frac{C_{s^*}(i)-1}{C_{s^*}(i)} + \frac{\lambda P_{H{s^*}}}{N_0\eta_{s^*}x_{s^*}(i)}\\
&=&\hspace{-3mm}\frac{1}{C_{s^*}(i)}\left(\frac{P_{Hs^*}A_{s^*}(i)}{\eta_{s^*}x_{s^*}(i)N_0}+B_{s^*}(i)\right).
\end{eqnarray}
\normalsize
\end{theorem}

\begin{IEEEproof}
Please refer to Appendix A.
\end{IEEEproof}

\vspace{-3mm}
\subsection{Constrained instantaneous BS transmit power}
Let us assume that, apart from the average power constraint $P_{avg}$, the instantaneous transmit power of the BS is limited to $P_M$. In this case, C3 of {\bf Q$_0$} is replaced by the following constraint:
\vspace{-2mm}
\begin{equation}
\label{cond}
\text{\^{C}}3: 0\leq p_0(i)\leq P_{M},\ \forall i,
\vspace{-2mm}
\end{equation}
leading to a new optimization problem:
\vspace{-1mm}
\begin{equation}
\text{{\bf Q$_1$}}\ \hspace{-1mm}:\hspace{-1mm}\text{ Identical to {\bf Q$_0$} with C3 replaced by \^{C}3}
\vspace{-1mm}
\end{equation}

The solution of problem {\bf Q$_1$}
is given in the following theorem:
\begin{theorem}\label{teorem2}
In epoch $i$, let us relabel the EHUs as
\vspace{-1mm}
\begin{eqnarray}
\notag
\label{sorting}
\frac{P_{H1}}{N_0\eta_1x_1(i)}\hspace{-3mm}&<&\hspace{-3mm}\frac{P_{H2}}{N_0\eta_2x_2(i)}<...<\frac{P_{Hg}}{N_0\eta_gx_g(i)}<P_M<\\
\hspace{10mm}& &\hspace{-13mm}<\frac{P_{Hg+1}}{N_0\eta_{g+1}x_{g+1}(i)}<...<\frac{P_{HK}}{N_0\eta_Kx_K(i)}.
\end{eqnarray}
\normalsize
If $g\geq s^*$, then Theorem 1 applies. Otherwise, let $m\in G_1=\{1,2,...,g\}$ and $n\in G_2=\{g+1,...,K\}$. In this case, the optimal BS transmit power is $p_0(i)=P_{M}$, and,  $\tau_m(i)$, $\tau_n(i)$, and $\tau_0(i)$ are obtained from (\ref{taum_EHS}), (\ref{taun_EHS}), and (\ref{tau_BS}), respectively, with $C_{s^*}(i)$ replaced by $C_g(i)$, where $C_g(i)$ is given by:
\vspace{-1mm}
\begin{equation}
\label{eq_Cg}
\log(C_g(i))-\frac{C_g(i)-1}{C_g(i)} + \lambda P_{M}=\frac{P_{M}A_g(i)+B_g(i)}{C_g(i)}.
\end{equation}
\end{theorem}

\begin{IEEEproof}
Due to the constrained space and the similarity with the proof of Theorem \ref{teorem1}, we give only a sketch of the proof. The problem is again solved using the dual Lagrangian method. For each epoch $i$, the new constraint \^{C}3 introduces another Lagrangian multiplier $\gamma(i)$, whose value is greater than 0 when $p_0(i)>P_{M}$. So, as long as $p_0(i)\leq P_{M}$, we have $\gamma(i)=0$ and the solution of {\bf Q$_1$} remains the same as the solution of {\bf Q$_0$}. When the solution of {\bf Q$_0$} is such that $p_0(i)> P_{M}$ then the right hand side of \^{C}3 is satisfied with equality, so $\gamma(i)>0$ and $p_0(i)$ becomes $p_0(i)=P_{M}$. In this case, the stations with $P_{Hk}/(N_0\eta_kx_k(i)) < P_{M}$ are in saturation and all other EHUs are in the linear regime, as seen from (\ref{sorting}). The optimal $\tau_0(i)$ and $\tau_k(i)$ are found following  \cite[Appendix]{lit21} while taking into account that some of the stations can be in the saturated regime.
\end{IEEEproof}
{\emph{Remark 1:}} If $P_{Hk} \to \infty, \forall k$ (corresponding to the linear EH model), Theorem 2 reduces to [2, Theorem 1]. In this case, all EHUs are in the linear regime in each epoch, and the optimal power allocation at the BS is binary (either zero or $P_M$).

\vspace{-3mm}
\section{Numerical Results}
\vspace{-2mm}
In this section, we assume a WPCN with Rayleigh fading channels, 
with $E[x_k'(i)] =10^{-3}\, D_k^{-3}$, where $D_k$ is the distance between the BS and EHU $k$. We assume a peak power constrained BS, and EHUs placed on a circle around the BS at a radius of 10 m.
For the proposed and the baseline resource allocation schemes, in our simulations, the harvested powers of all EHUs  are determined 
according to the practical EH characteristic curve in \cite[Fig. 5(b)]{stoo} obtained from measurements performed on EH circuits at 1 M$\Omega$ load. The proposed resource allocation scheme (denoted as "Proposed resource allocation") is based on Theorem 2. The parameters of the corresponding piece-wise linear EH model are obtained by fitting the two segments of the piece-wise curve to the measured characteristic in \cite[Fig. 5(b)]{stoo}, such that the slope of the linear part is determined by incident power levels lower than the value at which the power efficiency starts to decline rapidly ($-16$ dBm). This leads to $\eta_k = 0.2$ and $P_{Hk} = 9.2 \mu$W, $\forall k$.
We employ two baselines for performance comparison. Baseline 1 is a WPCN which employs the power and time allocation that is optimal for the linear EH model \cite[Theorem 1]{lit21} with $\eta_k = 0.2$.
Baseline 2 applies a constant power and time allocation scheme: in each epoch, the BS transmits with power $p_0(i) = P_M$, the EH phase duration is set to $\tau_0(i) = P_{avg}/P_M$ (such that C1 is satisfied with strict equality to ensure a fair comparison), and the IT phases are of equal duration $\tau_k(i) = (1-\tau_0(i))/K$, $\forall k$.

\begin{figure}[t!]
\vspace{0mm}
\begin{minipage}[b]{.45\linewidth}
\centering
\includegraphics[scale=0.43]{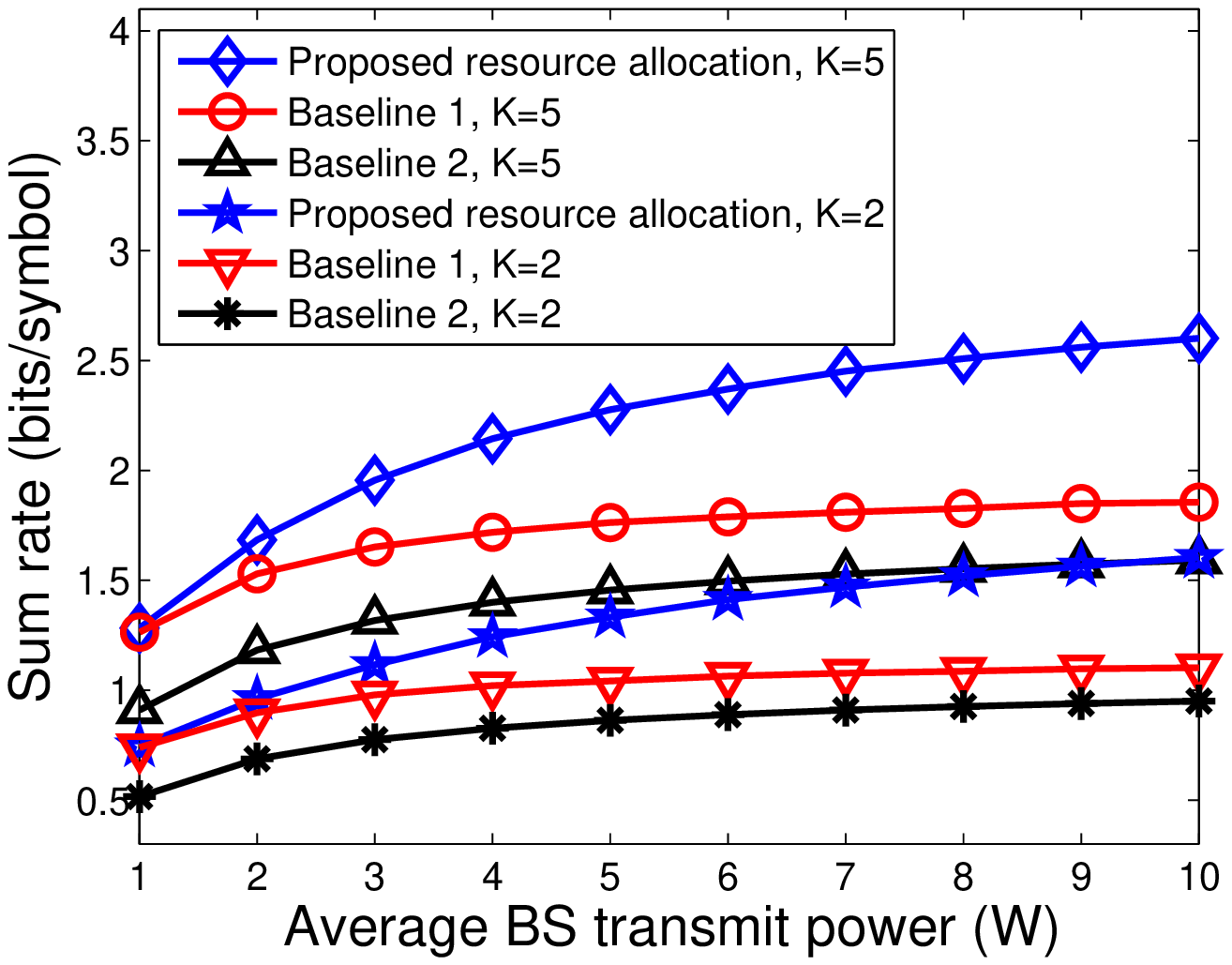}
\text{(a) $P_M=15P_{avg}$}
\end{minipage}
\begin{minipage}[b]{7cm}
\centering
\includegraphics[scale=0.43]{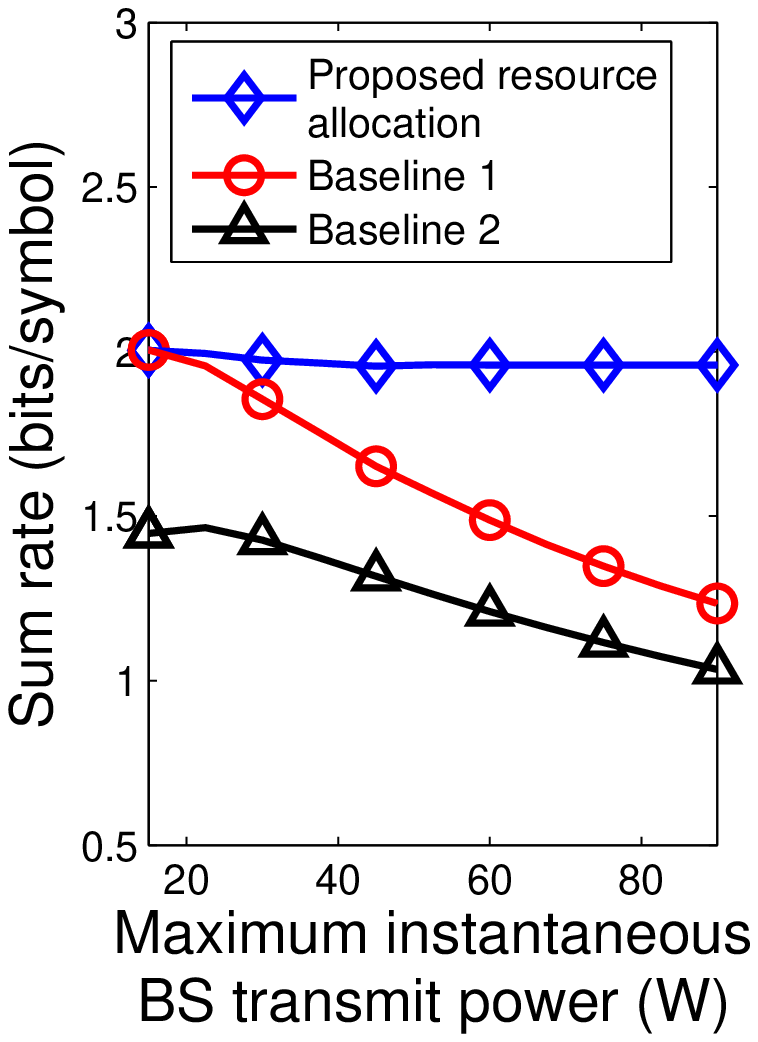}
\text{(b) $K=5$, $P_{avg}=3$ W}
\end{minipage}
\vspace{-7mm}
\caption{Sum rate as a function of a) $P_{avg}$ and b) $P_{M}$. } \label{fig1}
\end{figure}

Fig. \ref{fig1} shows the sum rates of the WPCNs obtained with different resource allocation schemes.
Fig. \ref{fig1} a) reveals that for small values of $P_{avg}$, the sum rate difference between the proposed resource allocation scheme 
and Baseline 1 is negligible, since for small values of $P_{avg}$ all EHUs operate in the linear regime for most of the time. As $P_{avg}$ increases, Baseline 1 uses the BS power less efficiently compared to the proposed resource allocation scheme, yielding lower sum rates.
Baseline 2 yields the worst results since no optimization is applied.
Fig. \ref{fig1} a) also reveals that a larger number of EHUs leads to a higher sum rate. This effect is caused by the larger amount of energy that can be harvested with more EHUs due to the broadcast nature of the wireless channel.

Fig. \ref{fig1} b) shows that increasing $P_M$ for a given $P_{avg}$  provides little benefit beyond a certain point, when the EH circuit  is non-linear.
As $P_M$ increases, the BS transmits less often with the instantaneous peak power, and inverts the channel of the EHUs instead.
On the contrary, when a linear EH model is assumed, the BS, if active, always transmits at power $P_M$  \cite{lit21}.
This results in a performance degradation for Baseline 1 as $P_M$ increases since the EHUs often become saturated resulting in a waste of power. A similar effect is observed for Baseline 2 since the BS also transmits at power $P_M$. Note that a small decrease  in the sum rate is observed even for the proposed model at $P_M = 35$ W. This is due to the mismatch between the actual non-linear EH characteristic used for simulations and the piece-wise linear EH model used for resource allocation design.

\vspace{-3mm}
\section{Conclusions}
\vspace{-2mm}
In this letter, we proposed a communication protocol that maximizes the uplink sum rate of WPCNs while taking into account the non-linearity of the EH RF-DC conversion through a piece-wise linear model. When the maximum instantaneous transmit power of the BS is not limited, the optimal protocol requires the BS to adapt its transmit power such that the harvested power of a certain EHU is exactly on the boundary between the linear regime and the saturated regime of the EHU's EH characteristic.
The proposed protocol outperforms the optimal protocol for a linear EH model since the latter forces the BS to transmit with maximum power causing a waste of power due to the saturation of the harvested power.

\vspace{-3mm}
\appendices
\section{Proof of Theorem 1}
\vspace{-2mm}
After substituting $e(i)=p_0(i)\tau_0(i)$ and representing the function $\min\{N_0x_k(i)e(i),P_{Hk}\tau_0(i)\}$ by its epigraph using the auxiliary variable $e_k(i)$, we obtain the following equivalent convex problem:

\vspace{-7mm}
\begin{eqnarray} \label{OP2}
\underset{\tau_{k}(i), \tau_{0}(i), e(i), e_k(i)} {\text{Maximize}} \ \frac{1}{M}\sum_{i=1}^M\sum_{k=1}^K\tau_{k}(i)\log\left(1+\frac{e_k(i)}{\tau_k(i)}{x}_{k}(i)\right)
\end{eqnarray}
\text{s.t.} 
\vspace{-5mm}
\begin{eqnarray}
\begin{array}{ll}
&\text{\={C}}1: \frac{1}{M}\sum_{i=1}^Me(i) \leq P_{avg} \\ \notag
&\text{\={C}}2: \tau_0(i)+\sum_{k=1}^K \tau_k(i) = 1, \forall i \\
&\text{\={C}}3: 0 \leq e(i), \forall i \\
&\text{\={C}}4: e_k(i)\leq N_0\eta_kx_k(i)e(i), \forall k,i\\
&\text{\={C}}5: e_k(i)\leq P_{Hk}\tau_0(i), \forall k,i \\
\end{array}
\end{eqnarray}

The Lagrangian of problem (\ref{OP2}) is given by:
\small
\begin{eqnarray}
\label{eq_lagr}
\notag
\mathcal{L} \hspace{-3mm}&=&\hspace{-3mm} \sum_{i=1}^M\sum_{k=1}^K\frac{\tau_{k}(i)}{M}\log\left(1+\frac{e_{k}(i)}{\tau_k(i)}{x}_{k}(i)\right)\hspace{-1mm}+\hspace{-1mm}\lambda \left(P_{avg} \hspace{-1mm}- \hspace{-1mm}\frac{1}{M}\sum_{i=1}^Me(i)\right)\hspace{-1mm}\\ \notag
\hspace{-5mm}&-&\hspace{-3mm} \sum_{i=1}^M\sum_{k=1}^K\beta_k(i)\left(e_k(i)- P_{Hk}\tau_0(i)\right)\hspace{-1mm}-\hspace{-1mm}\sum_{i=1}^M\varepsilon(i)\left(\tau_0(i)+\sum_{k=1}^K \tau_k(i)\hspace{-1mm}\right.\\
\hspace{-5mm}&-&\hspace{-3mm} \left.\hspace{-1mm} 1\vphantom{\frac{1}{2}}\right)- \sum_{i=1}^M\sum_{k=1}^K\alpha_k(i)(e_k(i)-N_0\eta_kx_k(i)e(i)) +\hspace{-1mm}\sum_{i=1}^M\mu(i)e(i)\hspace{-1mm}
\end{eqnarray}
\normalsize
where $\lambda$, $\varepsilon(i)$, $\mu_k(i)$, $\alpha_k(i)$, and $\beta_k(i)$ are the Lagrangian multipliers associated with \text{\={C}}1 - \text{\={C}}5, respectively. In epoch $i$, the slackness conditions must be satisfied for each $k$:
\vspace{-2mm}
\begin{eqnarray}
\label{slackness}
\notag
0\hspace{-3mm}&=&\hspace{-3mm}\mu(i)e(i)=\alpha_k(i)(e_k(i)-N_0\eta_kx_k(i)e(i))=\beta_k(i)(e_k(i)\\
\hspace{-3mm}&-&\hspace{-3mm} P_{Hk}\tau_0(i))=\varepsilon(i)\left(\tau_0(i)+\sum_{k=1}^K \tau_k(i) - 1\right).
\vspace{-3mm}
\end{eqnarray}
After differentiating $\mathcal{L}$ with respect to $\tau_k(i)$, $e_k(i)$, $e(i)$, and $\tau_0(i)$, and setting the result equal to zero, we get:
\small
\vspace{-2mm}
\begin{equation}
\label{der_tk}
\log\left(1+\frac{e_{k}(i)}{\tau_k(i)}{x}_{k}(i)\right)-\frac{\frac{e_{k}(i)}{\tau_k(i)}{x}_{k}(i)}{1+\frac{e_{k}(i)}{\tau_k(i)}{x}_{k}(i)}-\varepsilon(i)=0
\vspace{-2mm}
\end{equation}
\begin{equation}
\label{der_ek}
 -\alpha_k(i)-\beta_k(i)+{x}_{k}(i)/\left(1+\frac{e_{k}(i)}{\tau_k(i)}{x}_{k}(i)\right)=0
\vspace{-4mm}
\end{equation}
\begin{equation}
\label{der_e}
 -\lambda+\mu(i)+\sum_{k=1}^K\alpha_k(i)N_0\eta_kx_k(i)=0
\vspace{-2mm}
\end{equation}
\begin{equation}
\label{der_t0}
 -\varepsilon(i)+\sum_{k=1}^K\beta_k(i)P_{Hk}=0
\vspace{-2mm}
\end{equation}
\normalsize

From (\ref{der_tk}), we see that the BS receives the signals from all EHUs with the same signal-to-noise ratio (SNR) equal to $C(i) - 1$, i.e.,
\vspace{-2mm}
\begin{equation}
\label{pp_2}
e_{k}(i){x}_{k}(i)/\tau_k(i)=C(i)-1, 1\leq k\leq K,
\vspace{-2mm}
\end{equation}
where $C(i)$ satisfies $C(i)>1$, and from (\ref{der_ek}), we obtain:
\vspace{-2mm}
\begin{equation}
\label{pp_1}
x_k(i)/(\alpha_k(i)+\beta_k(i))=C(i), 1\leq k\leq K.
\vspace{-2mm}
\end{equation}

Now, assume that a single EHU, which is given index $s$, simultaneously satisfies \={C}4 and \={C}5  with equality.
Due to (\ref{assu}), the EHUs whose harvested power falls into the linear part of the EH model have indices $m\in S_1=\{s+1,s+2,\ldots,K\}$ and the EHUs whose harvested power falls into the saturation part have indices $n\in S_2=\{1,2,\ldots,s-1\}$. Due to  (\ref{slackness}), we have $\alpha_s(i)>0$, $\beta_s(i)>0$, $\alpha_m(i)>0$, $\beta_m(i)=0$, $\alpha_n(i)=0$, and $\beta_n(i)>0$. From (\ref{pp_1}), we therefore obtain
\begin{equation}
\label{eq_al_be}
\alpha_m(i)=\frac{x_m(i)}{C_s(i)},\ \beta_n(i)=\frac{x_n(i)}{C_s(i)}, \ \alpha_s(i)+\beta_s(i)=\frac{x_s(i)}{C_s(i)},
\vspace{-2mm}
\end{equation}
where $m\in S_1$ and $n\in S_2$.

{\bf{Case 1:}} Let us assume $e(i)>0$; thus $\mu(i)=0$.
For EHU $k = s$, we have $e(i)N_0\eta_sx_s(i)=P_{Hs}\tau_0(i)$, which yields: 
\begin{equation}
\label{pow_BSe}
p_0(i)=P_{Hs}/(N_0\eta_sx_s(i)).
\end{equation}

We now introduce an auxiliary variable $\rho_s(i)$ satisfying
$0<$ $\rho_s(i)<1$ that yields
$\alpha_s(i)=(1-\rho_s(i))\frac{x_s(i)}{C_s(i)}$ and $\beta_s(i)=$ $\rho_s(i)\frac{x_s(i)}{C_s(i)}$.
Based on (\ref{der_tk}), (\ref{der_e}), and (\ref{der_t0}), we obtain the following set of equations with  two unknowns ($\rho_s(i),C_s(i)$),
\begin{equation}
\label{wq_cs}
\log(C_s(i))-\frac{C_s(i)-1}{C_s(i)}=\frac{\rho_s(i) P_{Hs}x_s(i)+B_{s-1}(i)}{C_s(i)}
\end{equation}
\begin{equation}
\label{wq_cs_lambda}
{C_s}(i)=(A_{s}(i)+(1-\rho_s(i))N_0\eta_sx_s(i)^2)/\lambda
\end{equation}
\normalsize

There is exactly one solution of (\ref{wq_cs}), (\ref{wq_cs_lambda}), ($\rho_s^{*}(i),C_s^{*}(i)$), such that $0<\rho_s^{*}(i)<1$. To see this, we define an equation
\begin{equation}
\label{Fz}
f(z,a)=\log(z)-(z-1-a)/z
\end{equation}
which has root $z^* = (a-1)\left(W(e^{-1}(a-1))\right)^{-1}$ for $a>0$ and $z^*=1$ for $a=0$. From (\ref{wq_cs}), we note that $C_s(i)$ is an increasing function of $\rho_s(i)$ and from (\ref{wq_cs_lambda}), $C_s(i)$ is a linearly decreasing function of $\rho_s(i)$.
Thus, when setting $\rho = 0$ and $\rho = 1$, we respectively obtain: 
\begin{equation}
\label{eq_upper_lwer}
A_{s-1}(i)/\lambda>z_{s-1}\ \text{and} \ A_{s}(i)/\lambda<z_{s},
\end{equation}
where $z_{s-1}$ and $z_s$ satisfy $f(z_{s-1},B_{s-1}(i))=f(z_{s},B_{s}(i))=0$. Considering the definitions of $A_k(i)$ and $B_k(i)$, we can write $A_0(i)>A_1(i)>\ldots>A_K(i)=0$ and $0=B_0(i)<B_1(i)<\ldots<B_K(i)$, and thus, as $s$ increases from $0$ to $K$, the ratio $A_s(i)/\lambda$ decreases from $A_0(i)/\lambda$ to 0, whereas $z_s$ increases from $1$ to $z_K$. Therefore, assuming
$\frac{A_0(i)}{\lambda}>1$, equations (\ref{wq_cs}), (\ref{wq_cs_lambda}) have one solution for $0<\rho_s(i)<1$ which is found as in (\ref{s_value}) where $f(z_s,B_s(i))= 0$. Note that there is always a single solution for $s$ that satisfies (\ref{s_value}) because the regions $\left(\frac{A_{s-1}(i)}{\lambda}, \frac{A_{s}(i)}{\lambda}\right)$ and ($z_{s-1},z_s$) are adjacent to the corresponding regions of $s+1$.
 Once $s^*$ and $C_s^*(i)$ are determined, from (\ref{pp_2}), we obtain (\ref{taum_EHS}) and (\ref{taun_EHS}) and then, using \={C}2, where we substitute for $\tau_n(i)$ and $\tau_m(i)$, we obtain $\tau_0(i)$ as in (\ref{tau_BS}). The value of $\tau_0(i)$ can be used in (\ref{taum_EHS}) and (\ref{taun_EHS}) to calculate $\tau_m(i)$ and $\tau_n(i)$, respectively.

{\bf Case 2:} When $e(i)=0$, meaning that $p_0(i)=0$, we have $\mu(i)>0$, which yields the condition $\sum_{k=1}^{K}N_0\eta_kx_k^2(i)<\lambda$.


\end{document}